\documentclass[aps,twocolumn,prl,superscriptaddress,letterpaper]{revtex4-1}
\usepackage{amssymb}
\usepackage{mathtools}  
\usepackage[pdftex]{graphicx}
\usepackage[usenames, dvipsnames, svgnames, table]{xcolor}
\usepackage[colorlinks=true, citecolor=Blue,
            linkcolor=BrickRed, urlcolor=ForestGreen]{hyperref}
\usepackage{txfonts}
\usepackage{mathrsfs}
\usepackage{bm}

\begin{document}


\newcommand{\be}{\begin{equation}}
\newcommand{\ee}{\end{equation}}
\newcommand{\R}[1]{\textcolor{BrickRed}{#1}}
\newcommand{\B}[1]{\textcolor{blue}{#1}}


\title{Towards the Fundamental Quantum Limit of
Linear Measurements of Classical Signals}

\author{Haixing Miao}
\affiliation{School of Physics and Astronomy,    Institute of Gravitational Wave Astronomy,
University of Birmingham, Birmingham, B15 2TT, United Kingdom}
\author{Rana X Adhikari}
\affiliation{LIGO Laboratory, California Institute of Technology, Pasadena,
CA 91125, USA}
\author{Yiqiu Ma}
\affiliation{Theoretical Astrophysics 350-17, California Institute
of Technology, Pasadena,
CA 91125, USA}
\author{Belinda Pang}
\affiliation{Theoretical Astrophysics 350-17, California Institute
of Technology, Pasadena,
CA 91125, USA}
\author{Yanbei Chen}
\affiliation{Theoretical Astrophysics 350-17, California Institute
of Technology, Pasadena,
CA 91125, USA}


\begin{abstract}
The quantum Cram\'er-Rao bound (QCRB) sets a fundamental
limit for the measurement of classical signals with
detectors operating in the quantum regime.
Using linear-response theory and
the Heisenberg uncertainty relation, we
derive a general condition for achieving such a fundamental
limit. When applied to classical displacement measurements with a test mass, this condition leads to an explicit connection between
the QCRB and the Standard Quantum Limit which arises from a tradeoff between the measurement imprecision and quantum backaction; the QCRB
can be viewed as an outcome of a quantum non-demolition measurement with the backaction evaded.
Additionally, we show that the test mass is more a resource for improving measurement sensitivity than a victim of the quantum backaction, which suggests a new approach to enhancing the sensitivity of a broad class of sensors.
We illustrate these points
with laser interferometric gravitational wave
detectors.
\end{abstract}

\maketitle

{\it Introduction.---} In high-precision measurements of classical
signals, one challenge is to reduce various noise sources so that we can
measure the tiny change in the detector state caused by the signal.
This is often achieved by minimizing the coupling of the
detector to the environment. Eventually, we approach the quantum
regime with the dominant noise coming from the statistical
nature of the detector quantum state. Maximizing the
quantum-limited sensitivity requires proper preparation of the
detector state and measurements of its observables---a key
task in quantum metrology (cf. the review article by
Giovannetti {\it et al.}\,\cite{Giovannetti2011}). The quantum
Cram\'er-Rao bound (QCRB), derived in the pioneering works of
Helstrom\,\cite{Helstrom1967} and Holevo\,\cite{Holevo2011},
sets a fundamental limit to the maximum sensitivity
for a given detector state. As proved by Braunstein
{\it et al.}\,\cite{Braunstein1994, Braunstein1996},
this lower bound can be attained only if (i) the detector state
is pure and the right observable is measured, so that
the quantum Fisher information becomes equal to its classical
counterpart, and (ii) the estimator based upon
the measurement records is efficient, i.e., the mean squared
estimation error saturates the classical Cram\'er-Rao bound.

\begin{figure}[!b]
\includegraphics[width=\columnwidth]{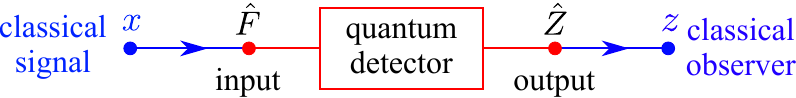}
\caption{(color online) A schematic for a quantum measurement of a classical signal
  using a linear detector. One degree of freedom of the detector
  is singled out as the input port, for which the observable,
  $\hat F$, is coupled to the signal, $x$, and another one as
  the output port with its observable, $\hat Z$, projectively
  measured by the observer. The detector is a quantum interface
  between two classical domains.
\label{fig:detector_model}}
\end{figure}

In linear measurements, as illustrated in Fig.\,\ref{fig:detector_model},
the detector input port observable, $\hat F$, is linearly coupled to the
signal, $x$. In the case of single-shot
detection of a single-parameter signal, this is modelled by the
interaction $\hat H_{\rm int}=-\,\hat F\,x\,\delta(t)$,
and the QCRB for the estimation error, $\sigma_{xx}$,
is (cf., Chapter 2 of Ref.\,\cite{Wiseman2010})
\begin{equation}\label{eq:QCRB_single-shot}
\sigma_{xx}^{\rm QCRB}=\frac{\hbar^2}{4 \langle\psi| \hat F^2
|\psi \rangle}\,,
\end{equation}
where $|\psi\rangle$ is the initial detector state, and we
assume that $\langle \psi|\hat F|\psi\rangle=0$. To attain it,
the output-port observable, $\hat Z$, that we measure needs
to satisfy~\cite{Braunstein1994},
\begin{equation}\label{eq:Braunstein_condition}
{\rm Re} [\langle \psi | \hat \Pi_z \, \hat F |\psi \rangle]
=0\quad \forall z\,,
\end{equation}
where ${\rm Re}[\cdot]$ means taking the real part, and the projection
operator $\hat \Pi_z$ is defined as
$\hat \Pi_{z}\equiv |z \rangle \langle z|$ with $|z\rangle$ being an
eigenstate of $\hat Z$ and $z$ the measurement outcome.
The maximum-likelihood
estimator of $x$, based upon $z$, will
be efficient if $|\psi\rangle$ is Gaussian,
or the sample size is
large~\footnote{ due to the central limit theorem. In this case, an extra factor of $1/N$ ($N$ the sample size) shall be included in the bound above.}.

For detecting signals with multi-dimensional parameters,
the QCRB is not as simple~\cite{Szczykulska2016b} as the one shown in
Eq.\,\eqref{eq:QCRB_single-shot}.
In particular, Tsang {\it et al.}\,\cite{Tsang2011} generalized
the QCRB to the linear measurement of a continuous signal $x(t)$
with an infinite-dimensional parameter space (specifically gravitational wave
detection using laser interferometers\,\cite{Caves81, Adhikari2014}).
For time-invariant, linear detectors with
$\hat H_{\rm int}=-\hat F\,x(t)$, they
showed that the QCRB for estimating the Fourier components, $x(\omega)$, of
the signal is
\begin{equation}\label{eq:QCRB_continuous}
\sigma_{xx}^{\rm QCRB}(\omega)=\frac{\hbar^2}
{4\bar S_{FF}(\omega)}\,,
\end{equation}
where $\bar S_{FF}$ is the symmetrized power spectral density
that describes the quantum fluctuations (uncertainty)
of $\hat F$.
Braginsky {\it et al.}\,\cite{Braginsky2000EQL} also derived a similar result, in terms of the signal-to-noise ratio.

Until now, it has not been shown generally how the QCRB in Eq.\,\eqref{eq:QCRB_continuous} can be achieved. This
is, however, crucial for applying the QCRB to guide the design of
quantum-limited linear sensors. We fill this gap by showing
general conditions for achieving the bound:
(1) the detector is at the quantum limit with minimum uncertainty,
and (2) the observables $\hat Z$ and $\hat F$ are uncorrelated
(in terms of cross-spectrum):
\begin{equation}\label{eq:condition1}
\bar S_{ZF}(\omega)=0\,.
\end{equation}
One can find the optimal $\hat Z$ satisfying the
second condition if the imaginary part of the input
susceptibility $\chi_{FF}$ vanishes:
\begin{equation}\label{eq:condition2}
{\rm Im}[\chi_{FF}(\omega)] = 0\,.
\end{equation}
When this is not the case and we only have the first condition satisfied,
the minimal estimation error will still be bounded:
\begin{equation}\label{eq:min_sigma_xx}
{\sigma_{xx}^{\rm QCRB}}\leq {\min \sigma_{xx}}\leq 2\,{\sigma_{xx}^{\rm QCRB}}.
\end{equation}
In deriving the above results, we use
the linear-response theory developed by Kubo\,\cite{Kubo1966},
which has previously been applied to analyze the quantum limited sensitivity
of linear detectors~\cite{Averin2003, Clerk2003, Braginsky92, Clerk2008}.
Additionally, we apply the recent result on the Heisenberg uncertainty
relation for continuous quantum measurements presented in Ref.~\cite{Miao2017}.

{\it Single-shot Measurements.---} Before discussing
the continuous measurements, we will first illustrate the basic
formalism using the example of a single-shot measurement with
$\hat H_{\rm int} = -\hat F \, x\,\delta(t)$.
Such an interaction will leave $\hat F$ unchanged, but induce a
shift on any observable that does not commute with $\hat F$.
Specifically, the solution to $\hat Z$ reads
\begin{equation}\label{eq:Z-single-shot}
\hat Z = \hat Z^{(0)} +
({i}/{\hbar})[\hat Z^{(0)},\,\hat F^{(0)}]\,x
\end{equation}
where the superscript $(0)$ denotes evolution under the detector free
Hamiltonian $\hat H_{\rm det}$. For linear detectors, the
canonical coordinates have classical-number  (i.e., not operator)
commutators, and $\hat H_{\rm det}$ only contains their linear or
quadratic functions. The relevant observables, $\hat Z$
and $\hat F$, also depend linearly on the canonical coordinates.
This justifies application
of linear-response theory, in which different quantities
are linked by classical-number susceptibilities. A brief introduction to
the linear-response theory is in the supplemental material.

\begin{figure}[!t]
  \includegraphics[width=0.9\columnwidth]{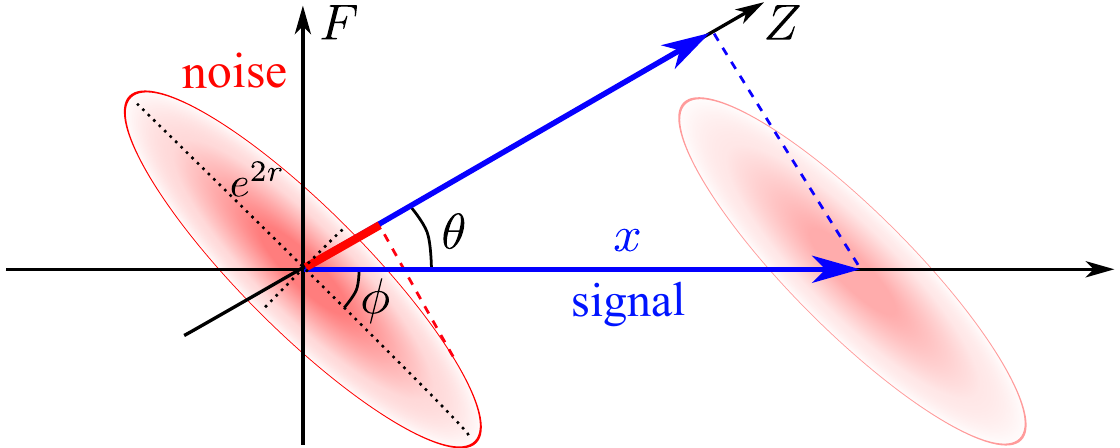}
    \caption{(color online) Illustration of the single-shot measurement with the detector
  in a pure, Gaussian, squeezed state (the noise ellipse represents its Wigner
  function). The optimal observable
  $\hat Z$ to achieve the QCRB is neither the conjugate variable of
  $\hat F$ (along the horizontal axis), which contains the largest signal,
  nor the one having the minimum noise (parallel to the semi-minor axis of
  the noise ellipse). Instead it is the one uncorrelated with
  $\hat F$ and $\tan\theta=\sin(2\phi)\sinh(2r)/[\cosh(2r)
  +\cos(2\phi)\sinh(2r)]$, in which $r$ and $\phi$ are the squeezing
  factor and angle,
  as derived in Ref.\,\cite{Braunstein1996}
  using Eq.\,\eqref{eq:Braunstein_condition}.}
\label{fig:condition-single-shot}
\end{figure}

In this example, we
introduce the following susceptibility:
\begin{equation}\label{eq:chiZF-single-shot}
\chi_{ZF}\equiv ({i}/{\hbar})[\hat Z^{(0)},\,\hat F^{(0)}]\,,
\end{equation}
which quantifies response of the detector output
to the signal: $\hat Z=\hat Z^{(0)}+\chi_{ZF}\,x$. Given the projective measurement
of $\hat Z$, we can construct an unbiased
estimator of the signal:
\begin{equation}\label{eq:xest-single-shot}
\hat x_{\rm est}=\hat Z/\chi_{ZF}\,.
\end{equation}
The resulting mean squared
error $\sigma_{xx}$ is determined by the quantum uncertainty of
$\hat Z^{(0)}$, i.e.,
\begin{equation}\label{eq:error-single-shot}
\sigma_{xx} \equiv {\rm Tr}[\hat \rho_{\rm det}
(\hat x_{\rm est}-x)^2]
= \sigma_{ZZ}/\chi_{ZF}^2\,,
\end{equation}
where $\sigma_{ZZ} \equiv {\rm Tr}[\hat \rho_{\rm det}
(\hat Z^{(0)})^2]$ assuming zero mean and $\hat \rho_{\rm det}$
is the density matrix of the detector initial state.
From the general Heisenberg uncertainty relation
between $\hat Z^{(0)}$ and $\hat F^{(0)}$:
\begin{equation}\label{eq:HUP-single-shot}
\sigma_{ZZ} \sigma_{FF} -\sigma_{ZF}^2 \ge ({\hbar^2}/{4})
\chi_{ZF}^2\,
\end{equation}
with $\sigma_{ZF} \equiv {\rm Tr}[\hat \rho_{\rm det}( \hat Z^{(0)}\hat F^{(0)}+
\hat F^{(0)}\hat Z^{(0)})/2]$ being their cross correlation, we obtain
\begin{equation}\label{eq:QCRB-single-shot-derivation}
\sigma_{xx} \ge \frac{\hbar^2}{4 \sigma_{FF}}+
\frac{\sigma_{ZF}^2}{\sigma_{FF}\chi_{ZF}^2}\ge
\frac{\hbar^2}{4 \sigma_{FF}}=
\sigma^{\rm QCRB}_{xx}\,.
\end{equation}
Achieving the QCRB therefore requires that the detector is
at quantum limit with minimum
uncertainty, i.e., in a
pure Gaussian state with Eq.\,\eqref{eq:HUP-single-shot} taking
the equal sign, and additionally
\begin{equation}\label{eq:condition-single-shot}
\sigma_{ZF}=0\,.
\end{equation}
Since $\hat Z=\int {\rm d}z \,\hat \Pi_{z}\, z$,
this condition is equivalent to
Eq.\,\eqref{eq:Braunstein_condition}. When discussing
a similar example, Braunstein {\it et al.}\,\cite{Braunstein1996}
derived the optimal $\hat Z$ using
Eq.\,\eqref{eq:Braunstein_condition}, as illustrated in
Fig.\,\ref{fig:condition-single-shot}.

{\it Continuous Measurements.---} The discussion for the continuous
measurements is quite similar to the single-shot case, but with
additional complications due to the involvement of many degrees
of freedom---the detector is a continuum field. We focus on linear
detectors that are time-invariant, i.e.,
having a time-independent $\hat H_{\rm det}$
and being in a stationary state $[\hat \rho_{\rm det},
\hat H_{\rm det}]=0$, allowing for frequency-domain
analysis of both dynamics and noise.

As in Eq.\,\eqref{eq:Z-single-shot},
$\hat Z$ in the continuous case is given by
\begin{equation}\label{eq:Z-continuous}
\hat Z(t)=\hat Z^{(0)}(t) + \int_{-\infty}^{\infty}{\rm d}t'
\chi_{ZF}(t-t') \,x(t')
\end{equation}
with the susceptibility,
$\chi_{ZF}\equiv (i/\hbar)[\hat Z^{(0)}(t),\, \hat F^{(0)}(t')]\Theta(t-t')$,
a function of the time
difference $t-t'$. In the frequency domain, it becomes
\begin{equation}\label{eq:z-continuous}
\hat Z(\omega) = \hat Z^{(0)}(\omega)+
\chi_{ZF}(\omega)x(\omega)\,,
\end{equation}
where $f(\omega)\equiv\int_{-\infty}^{+\infty} {\rm d}t\,
e^{i\omega t}f(t)$.
The unbiased estimator of $x(\omega)$, following
Eq.\,\eqref{eq:xest-single-shot}, is then
$\hat x_{\rm est}(\omega) = \hat Z(\omega)/{\chi_{ZF}(\omega)}$.

Given that the detector is in a stationary state, the quantum
fluctuation can be quantified by using the spectral density. There is
also a Heisenberg uncertainty relation
for the continuous measurements in terms of spectral
densities and susceptibilities (cf., Chapter VI in Ref.\,\cite{Braginsky92} or Ref.\,\cite{Miao2017}):
\begin{align}\nonumber
\bar S_{ZZ}(\omega)&\bar S_{FF}(\omega)-
|\bar S_{ZF}(\omega)|^2 \ge \frac{\hbar^2}{4}
|\chi_{ZF}(\omega)|^2 + \\
&\hbar \left |{\rm Im}[\bar S_{ZZ}(\omega)
\,\chi_{FF}(\omega)-\bar S_{ZF}^*(\omega)
\chi_{ZF}(\omega)]\right|
\,.
\label{eq:HUP-continuous}
\end{align}
Here the symmetrized spectral densities
$\bar S_{ZZ},\, \bar S_{FF}$ and $\bar S_{ZF}$ are defined
as $\bar S_{AB}(\omega)\equiv [S_{AB}(\omega)+S_{BA}(-\omega)]/2$
with the unsymmetrized one
$S_{AB}$ defined by $
{\rm Tr}[\hat \rho_{\rm det}\, \hat A^{(0)}
(\omega) \hat B^{(0)\dag}(\omega')]\equiv
2\pi \, S_{AB}(\omega)\delta(\omega-\omega')$~\cite{Clerk2008};
$\chi_{FF}$
is defined in the same way as $\chi_{ZF}$ in
Eq.\,\eqref{eq:Z-continuous} and with $\hat Z^{(0)}$
replaced by $\hat F^{(0)}$.

With Eq.\,\eqref{eq:HUP-continuous}, the  error
$\sigma_{xx}(\omega)\equiv{\bar S_{ZZ}(\omega)}
/{|\chi_{ZF}(\omega)|^2}$ for estimating $x(\omega)$
thus satisfies
\begin{equation}
\sigma_{xx}(\omega)
\ge \frac{\hbar^2}{4\bar S_{FF}}+
\frac{|\bar S_{ZF}|^2+\hbar  |{\rm Im}[\bar S_{ZZ}
\,\chi_{FF}-\bar S_{ZF}^*
\chi_{ZF}]|}{\bar S_{FF}|\chi_{ZF}|^2}\,.
\label{eq:QCRB-continuous-derivation}
\end{equation}
As proven in Ref.\,\cite{Miao2017},
when the detector is at the quantum limit,
i.e., in a pure, stationary,
Gaussian state---the multi-mode
squeezed state\,\cite{Blow1990}, not only does
Eq.\,\eqref{eq:HUP-continuous} become an equality,
but also we have
\begin{equation}\label{eq:chi_FF-equality}
{\rm Im}[\bar S_{ZZ}(\omega)
\,\chi_{FF}(\omega)-\bar S_{ZF}^*(\omega)
\chi_{ZF}(\omega)]|_{\rm quantum\; limit}=0\,.
\end{equation}
At this point, we only
require Eq.\,\eqref{eq:condition1}
to attain the QCRB---the first term in
Eq.\,\eqref{eq:QCRB-continuous-derivation}.

We now show that if Eq.\,\eqref{eq:condition2}
is satisfied, the optimal
observable $\hat Z$, which realizes Eq.\,\eqref{eq:condition1},
exists. In general, $\hat Z$
is a linear combination of two conjugate variables
(denoted by $\hat Z_{1,2}$) of the output port,
up to some constant:
\begin{equation}\label{eq:general-quadrature}
\hat Z(\omega)=\hat Z_1(\omega) \sin\theta
+ \hat Z_2(\omega)\cos\theta\,.
\end{equation}
Eq.\,\eqref{eq:condition1} can then be realized if there is
a real solution to $\theta$:
\begin{equation}\label{eq:quadrature-condition}
\tan\theta =-{\bar S_{Z_2 F}(\omega)}/
{\bar S_{Z_1F}(\omega)} \in {\rm Reals},
\end{equation}
or
${\rm Im}[\bar S_{Z_1 F}(\omega)\bar S_{Z_2F}^*(\omega)]=0$.
This turns out to be equivalent to
${\rm Im}[\chi_{FF}(\omega)]=0$ due to the following equality:
\begin{equation}\label{eq:readout-condition}
{\rm Im}[\bar S_{Z_1 F}(\omega)\bar S_{Z_2F}^*(\omega)]
=({\hbar}/{4}){\rm Im}[\chi_{FF}(\omega)]\,,
\end{equation}
which is generally valid for detectors at the quantum limit.

If ${\rm Im}[\chi_{FF}]$ is nonzero,
we will not find the optimal $\hat Z$ that exactly achieves the QCRB.
Nevertheless, the
estimation error $\sigma_{xx}$, minimized over all possible
$\theta$ in Eq.\,\eqref{eq:general-quadrature}, is still bounded as shown
in Eq.\,\eqref{eq:min_sigma_xx}. This is because
\begin{equation}\label{eq:nonzero_chi_FF}
\min_{\theta}\left|{\bar S_{ZF}(\omega)}/
{\chi_{ZF}(\omega)}\right|
\leq {\hbar}/{2}\,.
\end{equation}
Including
Eq.\,\eqref{eq:QCRB-continuous-derivation},
the above inequality implies Eq.\,\eqref{eq:min_sigma_xx}.
The detailed proofs for Eqs.\,\eqref{eq:readout-condition} and
\eqref{eq:nonzero_chi_FF} are provided in the supplemental
material\,\footnote{In addition to 
Refs.\,\cite{Kubo1966, Braginsky92,
BuCh2002,Clerk2008, Blow1990} mentioned in the main text, 
the supplemental material also 
includes Ref.\,\cite{Gardiner1985}.}.

{\it Classical Displacement Measurements.---}
The above discussion applies to general linear measurements. Here
we specifically look measurements of displacement;
the detector often consists of a quantum field and a test mass with
its position being displaced by a classical signal,
which can be a result of the action of a force signal.
The interaction between the field and the test mass leads to an important
sensitivity limit---the Standard Quantum Limit
(SQL), first derived by Braginsky\,\cite{Braginsky92}.
Below we show an explicit
connection between the SQL and the QCRB, and also
discuss the active role of the test mass
in enhancing the detector sensitivity.

In terms of a mathematical description, we denote the input
port observable of the field as $\hat {\cal F}$ and the output
as $\hat {\cal Z}$, to distinguish
from $\hat F$ and $\hat Z$ (relevant for the entire detector);
$\hat {\cal F}$ is coupled to
the test mass position $\hat q$ via
the interaction $-\hat q\,\hat{\cal F}$, and $\hat {\cal Z}$
is projectively measured. Solving the detector
dynamics leads to (in the frequency domain):
\begin{equation}\label{eq:field_detector_relation}
\hat F^{(0)}=\frac{\hat {\cal F}^{(0)}}{1-\chi_{qq}\chi_{\cal FF}}\,,\quad \hat Z^{(0)}=\hat {\cal Z}^{(0)}
+\frac{\chi_{\cal ZF}\chi_{qq}\hat{\cal F}^{(0)}}{1-\chi_{qq}\chi_{\cal FF}}\,.
\end{equation}
In the literature, the first term $\hat {\cal Z}^{(0)}$
of the detector output observable $\hat Z^{(0)}$ is referred to as the
imprecision noise; the second term, proportional to
 $\hat{\cal F}^{(0)}$, is the quantum backaction noise.

For the special case when
the input susceptibility of the field is zero: $\chi_{\cal FF}=0$,
the resulting estimation error is
\begin{equation}\label{eq:exp_error}
\sigma_{xx}=\frac{\bar S_{\cal ZZ}}{|\chi_{\cal ZF}|^2}+
2 {\rm Re}
\left[\chi_{qq}^*\frac{\bar S_{\cal ZF}}{\chi_{\cal ZF}}\right]
+|\chi_{qq}|^2 \bar S_{\cal FF}\,.
\end{equation}
If the imprecision noise
and the backaction noise are uncorrelated, i.e.
$\bar S_{\cal ZF}=0$, its lower bound will be the SQL:
\begin{equation}\label{eq:SQL}
\sigma_{xx}= \frac{\bar S_{\cal ZZ}}{|\chi_{\cal ZF}|^2}
+|\chi_{qq}|^2 \bar S_{\cal FF}\ge \hbar |\chi_{qq}|\equiv \sigma_{xx}^{\rm SQL}\,.
\end{equation}
The SQL can be surpassed by using quantum non-demolition (QND) measurements\,\cite{Braginsky1996}: e.g., coherent
noise cancellation schemes\,\cite{Tsang10}  or equivalently, optimal
readout schemes\,\cite{KLMTV2001} which cancel the
backaction noise. In particular, optimal readout schemes utilize
quantum correlations $\bar S_{\cal ZF}$, and
can be understood by applying the uncertainty relation
$\bar S_{\cal ZZ}\bar S_{\cal FF}\ge
|\bar S_{\cal ZF}|^2+\hbar^2 |\chi_{\cal ZF}|^2/4$
 to rewrite Eq.\,\eqref{eq:exp_error} as
\begin{equation}\label{eq:bound}
\sigma_{xx} \ge \frac{\hbar^2}{4 \bar S_{\cal FF}}+
\left|\frac{\bar S_{\cal ZF}}{\chi_{\cal ZF}}+\chi_{qq}\bar S_{\cal FF}\right|^2\ge \frac{\hbar^2}{4 \bar S_{\cal FF}}\,.
\end{equation}
The ultimate bound will be the QCRB if we read out the
optimal output observable satisfying $\bar S_{\cal ZF}/\chi_{\cal ZF}+\chi_{qq}\bar S_{\cal FF}=0$, which, from Eq.\,\eqref{eq:field_detector_relation}, is equivalent to Eq.\,\eqref{eq:condition1} shown earlier.
The SQL can therefore be viewed as arising from a
suboptimal readout scheme.

In cases where $\chi_{\cal FF}$ is not zero,
one can similarly show that the estimation error is again bounded by the QCRB: \begin{equation}\label{eq:QCRB_new}
\sigma_{xx}\ge \frac{\hbar^2}{4\bar S_{FF}}=\frac{\hbar^2}{4 \bar S_{\cal FF}}|1-\chi_{qq}\chi_{\cal FF}|^2\,.
\end{equation}
In contrast to Eq.\,\eqref{eq:bound}, here we have
a factor of  $|1-\chi_{qq}\chi_{\cal FF}|^2$,
which can be smaller than unity.
There are two equivalent interpretations:
(1) the test mass response is modified by the quantum field:
\begin{equation}\label{eq:chi_eff}
\chi_{qq}^{\rm eff}=\frac{\chi_{qq}}{1-\chi_{qq}\chi_{\cal FF}}
\,;
\end{equation}
and (2) the quantum fluctuations of the field are modified
by the test mass, as manifested by
the relation between $\hat F^{(0)}$ and $\hat{\cal F}^{(0)}$
in Eq.\,\eqref{eq:field_detector_relation}. The latter
highlights the active (and enhancing) role of the test mass,
rather than being a victim of the quantum backaction.
Below, we illustrate this using gravitational wave (GW)
detection with laser interferometers as an example.

\begin{figure}[!t]
  \includegraphics[width=\columnwidth]{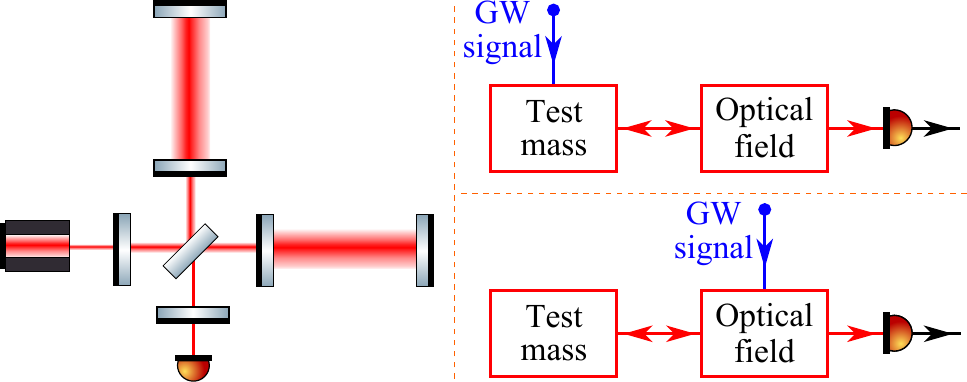}
  \caption{(color online) Schematic diagram of a LIGO-like interferometer (left).
  Two approximately equivalent physical pictures for the
  detection principle (right).}
  \label{fig:GW-detector}
\end{figure}

{\it Gravitational-wave Detection.---} A
typical GW detector,
such as LIGO\,\cite{AdvancedLIGO15}, is shown schematically in Fig.\,\ref{fig:GW-detector}. This is
an interferometer with Fabry-P\'erot arm cavities formed
by suspended mirrors (test masses).
The usual picture of the detection principle envisions
the GW as a tidal force on the test masses, and the resulting
differential motion being probed by the optical field. Another picture is to
view the GW as a strain directly coupled
to the optical field~\cite{Cooperstock1993, Tarabrin08}.
The latter is more appropriate when the GW wavelength
is comparable to or shorter than the interferometer arm
length, otherwise it is approximately equivalent to
the former. We will apply it in later discussions to highlight
the active role of the test mass mentioned earlier.

Putting the GW detection under the
general framework, the
classical signal is
\begin{equation}\label{eq:GW_signal}
x= L_{\rm arm}h_{\rm GW}\,,
\end{equation}
where $L_{\rm arm}$ is the arm length,
and $h_{\rm GW}$ is the GW strain.
The test mass motion that we care about is the differential mode
of the four mirrors in the two arms, with the susceptibility:
\begin{equation}\label{eq:sus_LIGO}
\chi_{qq}=-{4}/({M\, \omega^2})\,,
\end{equation}
where $M$ is the mirror mass.
The quantum field is the optical field,
coupled to the test mass via the radiation pressure.

As shown in Refs.\,\cite{scaling_law, Yanbei:review}, the
entire interferometer can be
mapped to a single-cavity-mode optomechanical device,
described by the standard cavity optomechanics\,\cite{Aspelmeyer2014}. The input observable $\hat{\cal F}$ is
 the time-varying part of the radiation pressure,
 which is proportional to the amplitude quadrature $\hat X$ of the cavity mode:
\begin{equation}\label{eq:F_LIGO}
\hat{\cal F}={2P_{\rm cav}}/{c}=\hbar g\hat X\,,
\end{equation}
of which the relevant susceptibility is given by~\cite{scaling_law}:
\begin{equation}\label{eq:chi_FF_exp}
\chi_{\cal FF}=
\frac{\hbar g^2\Delta}{(\omega-\Delta+i\gamma)
(\omega+\Delta+i\gamma)}\,.
\end{equation}
Here $g\equiv 2
\sqrt{\bar P_{\rm cav }\omega_{\rm cav} /
(\hbar L_{\rm arm} c )}$
with $\bar P_{\rm cav}$ the average optical power inside
 the cavity and $\omega_{\rm cav}$ the
 cavity resonant frequency; $\Delta=\omega_0-\omega_{\rm cav}$ is the
detuning of the laser frequency $\omega_0$; $\gamma$ is the cavity
bandwidth. The output observable $\hat {\cal Z}$
is a linear combination of the amplitude and phase quadrature of
the outgoing field at the dark (differential) port.

\begin{figure}[!t]
  \includegraphics[width=\columnwidth]{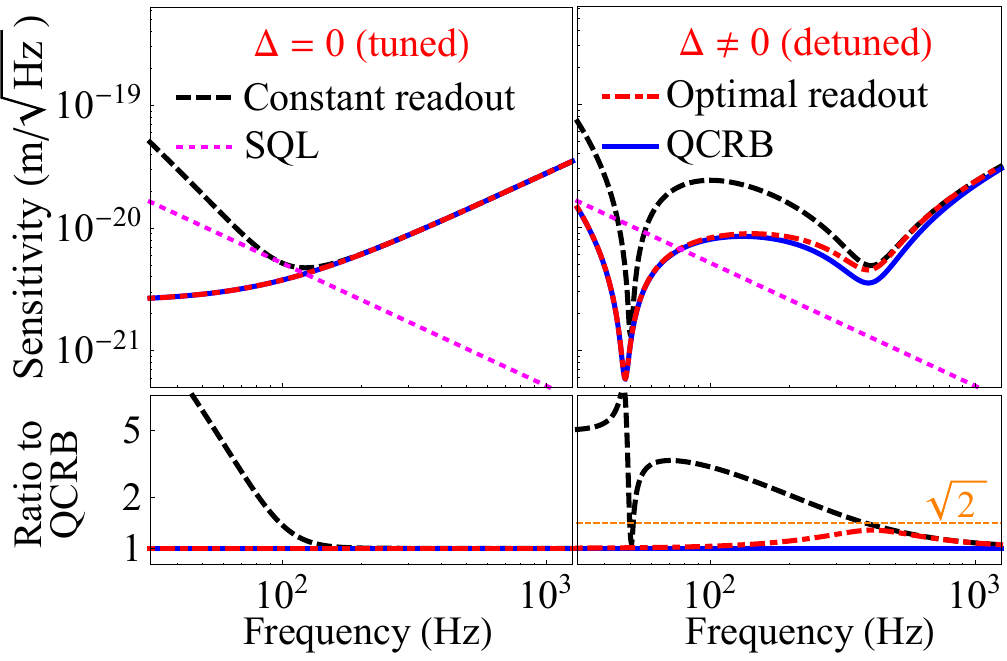}
  \caption{(color online) The top row
shows the QCRB (solid curve) for LIGO-type
GW detector with detuning frequency $\Delta =0$ (left) and
 $\Delta/(2\pi)=400$\,Hz (right), and various sensitivity curves for
 comparison: (i) dash curve---constant phase quadrature readout,
(ii) dash-dot curve---readout quadrature optimized
 to maximize sensitivity at each frequency, and
(iii) dot curve---the SQL $\sqrt{4\hbar/(M\omega^2)}$.
The bottom row shows the ratio to the QCRB for selected curves.
Other relevant parameters are: $M = 40$\,kg, $P_{\rm cav}=800$\,kW,
$L_{\rm arm}=4$\,km, $\gamma/(2\pi)\approx 100$\,Hz,
and laser frequency $\omega_0/(2\pi)\approx 3\times 10^{14}$\,Hz.}
   \label{fig:comparison}
\end{figure}

In Fig.\,\ref{fig:comparison}, we plot
the resulting QCRB for the two cases: $\Delta=0$ (tuned) and
$\Delta\neq 0$ (detuned), assuming
other parameters similar to LIGO. In comparison, we have also
included the SQL, and the estimation error $\sigma_{xx}^{1/2}$,
i.e. the sensitivity, for the phase
quadrature readout and the optimal readout.
The tuned case having $\chi_{\cal FF}=0$ provides a concrete example of Eq.\,\eqref{eq:SQL} and Eq.\,\eqref{eq:bound}. Indeed,
the optimal readout, which surpasses the SQL by
canceling the backaction noise, leads to
a sensitivity exactly equal to the QCRB.

In the detuned case with $\chi_{\cal FF}\neq0$, the first point we want to highlight is that
the maximum difference between the optimal-readout
sensitivity, considered by Harms {\it et al.}\,\cite{Harms2003a},
and the QCRB is at most $\sqrt{2}$ in amplitude,
in accordance with our general result Eq.\,\eqref{eq:min_sigma_xx}.
The second
point is that there are two noticeable dips in the QCRB. They
imply that
the amplitude
quadrature of the cavity mode has higher fluctuations
around these dips than other frequencies. Both can be interpreted as arising from
positive feedback induced optical resonance.
The higher frequency one coincides with the detuning frequency,
which is at the cavity resonance.
The low frequency one provides an example of
the extra factor $|1-\chi_{qq}
\chi_{\cal FF}|^2$ in Eq.\,\eqref{eq:QCRB_new}. Physically, this
has to do with
the ponderomotive squeezing (or amplification)
effect\,\cite{KLMTV2001, Corbitt2006}, which
recently has been demonstrated experimentally\,\cite{Brooks2012,
Purdy2013, Safavi-Naeini2013a}. The test mass acts as a
Kerr-type nonlinear medium converting the amplitude
fluctuations into the phase fluctuations, which  in turn,
feeds back to the amplitude quadrature due to the cavity detuning.
Since the test mass susceptibility goes as
$1/\omega^{2}$, cf. Eq.\,\eqref{eq:sus_LIGO}, the feedback gain is
frequency dependent, resulting in the sharp resonance feature.
The underlying physics is similar to the intra-cavity squeezing
studied theoretically by Peano {\it et al.}~\cite{Peano2015}
 and experimentally by Korobko {\it et al.}~\cite{Korobko2017}.

An equivalent interpretation of the low frequency dip
was presented in Refs.~\cite{BuCh2002, scaling_law}.
It was attributed to the so-called optical spring effect,
an example of Eq.\,\eqref{eq:chi_eff}---the
optomechanical interaction changes the test mass dynamics by
creating a new mechanical resonance, around
which the response to GWs is enhanced. The previous optical
feedback interpretation, however, removes the distinction
between optics and mechanics---the role of the latter also
modifies the quantum fluctuations of the optical field.
This suggests a new approach to designing
optomechanical sensors. We can add proper optical
filters in the feedback loop, together with the internal
ponderomotive
squeezing, to shape the optical feedback gain, so that
the quantum fluctuation of the field is enhanced in
the frequency band of interest. Since the sensitivity
using the optimal readout is
bounded, cf. Eq.\,\eqref{eq:min_sigma_xx},
this will result in high detector sensitivity at relevant
frequencies, with limitations only coming from the losses.
Incorporating the effect of losses is critical and the subject
of future work.

{\it Acknowledgements.---} We would like to thank members
of the LSC MQM, AIC, and QN groups for fruitful discussions.
HM is supported by UK STFC Ernest Rutherford Fellowship (Grant
No. ST/M005844/11). RXA is supported by NSF grant PHY-0757058.
YM, BP, and YC are supported by NSF PHY-0555406,
PHY-0653653, PHY-0601459, PHY-0956189, PHY-1068881, as
well as the David and Barbara Groce startup fund at Caltech.
RXA, BP, and YC gratefully acknowledge funding provided by
the Institute for Quantum Information and Matter, an NSF
Physics Frontier Center with support of the Gordon and
Betty Moore Foundation.

%



\pagebreak
\clearpage
\begin{center}
\textbf{\large Supplemental Material}
\end{center}
\setcounter{equation}{0}
\setcounter{figure}{0}
\setcounter{table}{0}
\makeatletter
\renewcommand{\theequation}{S\arabic{equation}}
\renewcommand{\thefigure}{S\arabic{figure}}
\renewcommand{\bibnumfmt}[1]{[S#1]}
\renewcommand{\citenumfont}[1]{S#1}

\section{I. Linear-response theory}

Here we briefly introduce the linear-response theory
that has been applied in our analysis.
One can refer to Refs.\,\cite{Kubo1966s, Braginsky92s,
BuCh2002s,Clerk2008s}
for more details. Given the model illustrated in
Fig.\,1 of the main paper, the
Hamiltonian for the measurement setup is
\begin{equation}\label{eq:Hamiltonian}
\hat H_{\rm tot}= \hat H_{\rm det}+\hat H_{\rm int}\,,
\end{equation}
where $\hat H_{\rm det}$ is the free Hamiltonian for
the detector,
and $\hat H_{\rm int}$ describes
the coupling between the classical signal and
the detector. We consider the
steady state with the coupling turned on
at $t=-\infty$. The solution to any operator
$\hat A$ of the detector at
time $t$ in the Heisenberg picture is given by
\begin{equation}\label{eq:solution}
\hat A(t) = \hat U_I^{\dag}(-\infty, t) \hat A^{(0)}(t)
\hat U_I(-\infty, t)\,
\end{equation}
with $\hat A^{(0)}(t)$ denoting the operator under the free
evolution:
\begin{equation}\label{eq:A_free}
\hat A^{(0)}(t)\equiv
\hat U^{\dag}_0(-\infty, t)\hat A\,\hat U_0(-\infty, t)\,.
\end{equation}
The unitary operator for the free-evolution part
is defined as $\hat U_0(-\infty, t)\equiv {\cal T}\exp
\{-(i/\hbar)\int_{-\infty}^t {\rm d}t'\hat H_{\rm det}(t')\}$
with $\cal T$ being the time-ordering, and, for the interaction
part, we have defined $\hat U_I(-\infty, t)\equiv {\cal T}
\exp\{-(i/\hbar)\int_{-\infty}^t {\rm d}t' \hat
H_{\rm int}^{(0)}(t')\}$.

For the measurement to be linear, $\hat H_{\rm det}$
only involves linear or quadratic functions of canonical
coordinates, among which their
commutators are classical numbers, i.e., not
operators; the
interaction $\hat H_{\rm int}$ is in the
bilinear form:
\begin{equation}\label{eq:interaction_Hamiltonian}
\hat H_{\rm int}=- \hat Fx(t)\,.
\end{equation}
As a result, Eq.\,\eqref{eq:solution} leads to
the following
exact solution to the input-port observable $\hat F$
and output-port observable $\hat Z$:
\begin{align}
\label{eq:solution_Z}
\hat Z(t) & =\hat Z^{(0)}(t) +
\int_{-\infty}^{+\infty} {\rm d}t' \chi_{ZF}(t,\, t') \, x(t')\,,\\
\label{eq:solution_F}
\hat F(t) & =\hat F^{(0)}(t)+\int_{-\infty}^{+\infty} {\rm d}t'
\chi_{FF}(t,\, t') \, x(t')\,.
\end{align}
The susceptibility
$\chi_{AB}$ ($A, B = Z, F$), which describes the detector
response to the signal, is defined as
\begin{equation}\label{eq:susceptibility_definition}
\chi_{AB}(t,\, t')\equiv \frac{i}{\hbar}[\hat A^{(0)}(t),\,
\hat B^{(0)}(t')]\Theta(t-t')
\end{equation}
with $\Theta(t)$ being the Heaviside function.
Notice that the susceptibilities are classical
numbers and only involve operators under the free
evolution,
which are consequences of the detector being
linear.

For the measurement to be continuous, we need to be able
to projectively measure the output-port observable at
different times precisely without introducing additional
noise. This can happen only if $\hat Z(t)$ commutes with itself
at different times, namely,
\begin{equation}\label{eq:meas_cond}
 [\hat Z(t),\,\hat Z(t')]=0 \quad \forall t, t'\,.
\end{equation}
It is called the condition of
simultaneous measurability in Ref.\,\cite{BuCh2002s} which also
shows that it implies
\begin{equation}
 [\hat Z^{(0)}(t),\,\hat Z^{(0)}(t')] = [\hat F^{(0)}(t),\,\hat Z^{(0)}(t')] \Theta(t-t')=0\,,
\end{equation}
or equivalently,
\begin{equation}\label{eq:chi_zz_chi_fz}
\chi_{ZZ}(t,\, t')=\chi_{FZ}(t,\, t')=0\,,
\end{equation}
which is central to the discussion of continuous, linear
quantum measurements.

When the free Hamiltonian for the detector is
time-independent, the susceptibility will only depend
on the time difference, i.e.,
\begin{equation}
\chi_{AB}(t,\, t')=\chi_{AB}(t-t')\,,
\end{equation}
which is the case considered in the main paper.
This allows us to move into the frequency domain, and
rewrite Eqs.\,\eqref{eq:solution_Z} and \eqref{eq:solution_F} as
\begin{align}
\label{eq:solution_Z_freq}
\hat Z(\omega) & =\hat Z^{(0)}(\omega) +
\chi_{ZF}(\omega) \, x(\omega)\,,\\
\label{eq:solution_F_freq}
\hat F(\omega) & =\hat F^{(0)}(\omega)+
\chi_{FF}(\omega) \, x(\omega)\,.
\end{align}
in which the Fourier transform $\hat A(\omega)\equiv
\int_{-\infty}^{+\infty}{\rm d}t\, e^{i\omega t}\hat A(t)$.
Furthermore, we consider the detector being in a stationary state, i.e.,
its density matrix $\hat \rho_{\rm det}$ commuting with
$\hat H_{\rm det}$. The statistical property
of the relevant operators, which defines the quantum
noise of the detector, can then be quantified
by using the frequency-domain spectral density, which is given by
\begin{equation}\label{eq:spectral_unsym}
S_{AB}(\omega)\equiv \int_{-\infty}^{+\infty}
{\rm d}t\, e^{i\omega t} {\rm Tr}[\hat \rho_{\rm det}\, \hat A^{(0)}(t+\tau) \hat B^{(0)}(\tau)]\,,
\end{equation}
where $\tau$ can be arbitrary due to the stationarity, and we have assumed ${\rm Tr}[\hat \rho_{\rm det}\hat A]={\rm Tr}[\hat \rho_{\rm det}\hat B]=0$ without loss of generality. Or
equivalently, the spectral density can also be defined through
\begin{equation}\label{eq:spectral_unsym}
{\rm Tr}[\hat \rho_{\rm det}\, \hat A^{(0)}(\omega)
\hat B^{(0)\dag}(\omega')]\equiv 2\pi\,S_{AB}(\omega)
\delta(\omega-\omega')\,.
\end{equation}
The corresponding symmetrized version of the previously
defined spectral density
is
\begin{equation}\label{eq:spectral_sym}
\bar S_{AB}(\omega) \equiv \frac{1}{2}[S_{AB}(\omega)
+S_{BA}(-\omega)]\,,
\end{equation}
which is a summation of both the
positive-frequency and negative-frequency
spectra.

From the definitions of the susceptibility and spectral
density, we have a general equality relating
them to each other:
\begin{equation}\label{eq:sus_spec_relation}
\chi_{AB}(\omega)-\chi^*_{BA}(\omega)=
\frac{i}{\hbar}[S_{AB}(\omega)-S_{BA}(-\omega)]\,.
\end{equation}
When applying this to the case with $\hat A=\hat B$, it
 leads to the famous Kubo's formula:
\begin{equation}\label{eq:kubo}
{\rm Im}[\chi_{AA}(\omega)]=
\frac{1}{2\hbar}[S_{AA}(\omega)-S_{AA}(-\omega)]\,.
\end{equation}
Such an imaginary part of the susceptibility
${\rm Im}[\chi_{AA}(\omega)]$
quantifies the
dissipation, and, in the thermal
equilibrium, it is related to the symmetrized
spectral density $\bar S_{AA}(\omega)$
through the fluctuation-dissipation theorem. The
measurement process is
far from the thermal equilibrium,
and therefore the usual fluctuation-dissipation theorem cannot
be applied.
Nevertheless, when the detector is ideal at the quantum
limit with minimum uncertainty,
we can also find some general relations between the
susceptibility and the symmetrized spectral density,
e.g., Eq.\,(18) and Eq.\,(21) in the main paper, the later
of which will be proven in the next section.

\section{II. Proof of Eq.\,(21) }

Here we show the proof of Eq.\,(21) in the
main paper. In the continuous, linear
measurements, the detector is a continuum field
that contains many degrees of freedom
which are coupled to each other through
the free evolution. The degrees of freedom for
the input and output port that we pick are
continuously driven by the ingoing part of
the continuum field, which is similar to the
in field introduced in Ref.\,\cite{Gardiner1985s}.
In the steady state with the initial
condition decaying away, their observables
$\hat Z_{1, 2}$ and
$\hat F$ can be generally represented in terms
of the ingoing field:
\begin{align}\label{eq:Z_F_d1}
\hat Z^{(0)}_{1,2}(t) &= \int_{-\infty}^{\infty}{\rm d}t'{\mathscr Z}_{1,2}(t-t')
\hat d(t')+ {\rm h.c.}\,,\\\label{eq:Z_F_d2}
\hat F^{(0)}(t)  &= \int_{-\infty}^{\infty}{\rm d}t'{\mathscr F}(t-t')\hat d(t') +
{\rm h.c.}\,.
\end{align}
Here ${\mathscr Z}$ and ${\mathscr F}$ are some
complex-valued functions;
$\rm h.c.$ denotes Hermitian conjugate;
$\hat d(t)$ is annihilation operator of the ingoing field
that satisfies the following
commutator relation:
\begin{equation}\label{eq:d_comm}
[\hat d(t),\,\hat d^{\dag}(t')]=\delta(t-t')\,.
\end{equation}
In the frequency domain, Eqs.\,\eqref{eq:Z_F_d1} and
\eqref{eq:Z_F_d2} can be rewritten as
\begin{align}\label{eq:general-io}
\hat Z^{(0)}_{1,2}(\omega) &= {\mathscr Z}_{1,2}(\omega)
\hat d(\omega)+
{\mathscr Z}_{1,2}^*(-\omega)
\hat d^{\dag}(-\omega)\,,\\
\label{eq:Fmode}
\hat F^{(0)}(\omega)  &= {\mathscr F}(\omega)\hat d(\omega) +
{\mathscr F}^*(-\omega)\hat d^{\dag}(-\omega)\,,
\end{align}
and the commutator for the ingoing field is
\begin{equation}\label{eq:comm_d_mode}
[\hat d(\omega),\, \hat d^{\dag}(\omega')]=2\pi\,
\delta(\omega-\omega')\,.
\end{equation}

A natural choice for the output port is the outgoing part
of the continuum field,
similar to the out field in Ref.\,\cite{Gardiner1985s},
which guarantees that the condition in
Eq.\,\eqref{eq:meas_cond} can be fulfilled due to
causality. Its two conjugate
variables $\hat Z_{1,2}$ satisfies
\begin{equation}
[\hat Z_k(t),\, \hat Z_l(t')]
= -\sigma^{kl}_y\delta(t-t')\,,
\end{equation}
where $k, l = 1,2 $ and $\sigma_y$ is the Pauli matrix.
In the frequency domain, the above commutator reads
\begin{equation}\label{eq:comm_z}
[\hat Z_k(\omega),\, \hat Z_l^{\dag}(\omega')]
= -2\pi\,\sigma_y^{kl}\delta(\omega-\omega')\,.
\end{equation}
Together with Eq.\,\eqref{eq:comm_d_mode},
this implies the following constraint on those functions in
Eq.\,\eqref{eq:general-io}:
\begin{equation}\label{eq:zk-constraints}
{\mathscr Z}_k(\omega){\mathscr Z}^*_l(\omega)-
{\mathscr Z}_k^*(-\omega){\mathscr Z}_l(-\omega)
= - \sigma_y^{kl}\,,
\end{equation}
which is an important equality for the proof.

We first prove Eq.\,(21) in the case
when the detector is in the vacuum state, i.e.,
\begin{equation}\label{eq:vacc}
\hat \rho_{\rm det}= |0\rangle \langle 0|\,.
\end{equation}
Correspondingly, we have $
{\rm Tr}[\hat \rho_{\rm det}\, \hat d(\omega)
\hat d^{\dag}(\omega')] =2\pi \,\delta(\omega-\omega')$ and
${\rm Tr}[\hat \rho_{\rm det}\, \hat d^{\dag}(\omega)
\hat d(\omega')] =0$,
which are equivalent to
\begin{align}
S_{\hat d\hat d^{\dag}}(\omega)=1\,,\quad
S_{\hat d^{\dag}\hat d}(\omega)=0\,.
\end{align}
From Eqs.\,\eqref{eq:general-io} and \eqref{eq:Fmode},
the above spectral density for $\hat d$ leads to
\begin{align}\label{eq:spec_zkF}
S_{Z_{1,2} F}(\omega) &={\mathscr Z}_{1,2}(\omega) {\mathscr F}^*(\omega)\,,\\
\label{eq:spec_FF}
S_{FF}(\omega) &= |{\mathscr F}(\omega)|^2\,.
\end{align}
Using the constraint in Eq.\,\eqref{eq:zk-constraints}
and the definition of symmetrized spectral density
 Eq.\,\eqref{eq:spectral_sym}, we find
\begin{equation}\label{eq:readout-condition}
{\rm Im}[\bar S_{Z_1 F}(\omega)\bar S_{Z_2F}^*(\omega)]
= \frac{1}{8}[S_{FF}(\omega)-S_{FF}(-\omega)]\,.
\end{equation}
With the Kubo's formula Eq.\,\eqref{eq:kubo}:
\begin{equation}
{\rm Im}[\chi_{FF}(\omega)]=
\frac{1}{2\hbar}[S_{FF}(\omega)-S_{FF}(-\omega)]\,,
\end{equation}
finally it gives rise to Eq.\,(21) in the main paper, i.e.,
\begin{equation}\label{eq:zF_FF_relation}
 {\rm Im}[\bar S_{Z_1 F}(\omega)\bar S_{Z_2F}^*(\omega)] =
 \frac{\hbar}{4}{\rm Im}[\chi_{FF}(\omega)]\,.
\end{equation}

We can further show that
Eq.\,\eqref{eq:zF_FF_relation} also holds
for the general, stationary, pure Gaussian state---multi-mode
squeezed state $\hat \rho_{\rm det}= \hat {\cal S}|0\rangle\langle 0|
 \hat {\cal S}^{\dag}$, in which the
squeezing operator $\hat {\cal S}$ is defined as\,\cite{Blow1990s}
\begin{equation}\label{eq:sqz_state}
\hat {\cal S}\equiv
\exp\left\{\int_{-\infty}^{\infty} \frac{{\rm d}\omega}{2\pi}
 [\xi(\omega) \hat d^{\dag}(\omega)
 \hat d^{\dag}(-\omega)-{\rm h.c.}]\right\}\,
\end{equation}
with $\xi(\omega)=\xi(-\omega)$.
This is because
$\hat {\cal S}$ only makes a Bogoliubov transformation
of $\hat d$. The spectral densities in
Eqs.\,\eqref{eq:spec_zkF} and \eqref{eq:spec_FF} are in the same form
as in the case of vacuum state,
after replacing ${\mathscr Z}_{1,2}$ by ${\mathscr Z}'_{1,2}$
and ${\mathscr F}$ by ${\mathscr F}'$:
\begin{align}
{\mathscr Z}_{1,2}'(\omega)&\equiv {\mathscr Z}_{1,2}(\omega)\cosh r_s+
e^{-i\phi_s}{\mathscr Z}_{1,2}^*(-\omega)\sinh r_s\,,\\
{\mathscr F}'(\omega)&\equiv {\mathscr F}(\omega)\cosh r_s+
e^{-i\phi_s}{\mathscr F}^*(-\omega)\sinh r_s\,,
\end{align}
where the real-valued functions $r_s$ and $\phi_s$ are
 defined through $\xi(\omega)\equiv r_s(\omega)e^{i\phi(\omega)}$.
Such a transform will leave Eq.\,\eqref{eq:zF_FF_relation} unchanged.

\section{III. Minimum of $|\bar S_{ZF}/\chi_{ZF}|$}

Here we prove Eq.\,(22) of the main paper. Given the output-port
observable
$\hat Z=\hat Z_1\sin\theta+\hat Z_2\cos\theta$, we have
\begin{align}
 \bar S_{ZF}(\omega)&=
 \bar S_{Z_1F}(\omega)\sin\theta
 +\bar S_{Z_2F}(\omega)\cos\theta\,, \\
 \chi_{ZF}(\omega)&=\chi_{Z_1F}(\omega)
 \sin\theta+\chi_{Z_2F}(\omega)\cos\theta\,.
\end{align}
The absolute value of their ratio is simply, for $\theta\neq 0$,
\begin{equation}
{\cal R}\equiv \left|\frac{\bar S_{ZF}(\omega)}
{\chi_{ZF}(\omega)}\right|=
 \left|\frac{\bar S_{Z_1F}(\omega)
 +\bar S_{Z_2F}(\omega)\cot\theta}
 {\chi_{Z_1F}(\omega)
 +\chi_{Z_2F}(\omega)\cot\theta}\right|\,.
\end{equation}

Using Eqs.\,\eqref{eq:chi_zz_chi_fz} and
\eqref{eq:sus_spec_relation}, we can express the susceptibility
$\chi_{Z_{1,2}F}$ in terms of the unsymmetrized spectral density:
\begin{equation}
 \chi_{Z_{1,2}F}(\omega)=\frac{i}{\hbar}[S_{Z_{1,2} F}(\omega)-
 S_{F Z_{1,2}}(-\omega)]\,.
\end{equation}
Form the expressions for $S_{Z_{1,2} F}$ shown
in Eq.\,
\eqref{eq:spec_zkF}, the above ratio can be rewritten as
\begin{equation}
{\cal R}=\frac{\hbar}{2}\left|\frac{1+\alpha\beta}
{1-\alpha\beta}\right|\,,
\end{equation}
where we have defined
\begin{align}
 \alpha&\equiv\frac{{\mathscr Z}_1^*(-\omega)+
 {\mathscr Z}_2^*(-\omega)\cot\theta}{{\mathscr Z}_1(\omega)+
 {\mathscr Z}_2(\omega)\cot\theta}\,,\\\label{eq:beta}
 \beta&\equiv \frac{{\mathscr F}(-\omega)}
 {{\mathscr F}^*(\omega)}\,.
\end{align}
With the constraint Eq.\,\eqref{eq:zk-constraints},
one can show that
\begin{equation}
|\alpha|=1\,.
\end{equation}
We can therefore write $\alpha$ as $e^{i \phi_{\alpha}}$ with $\phi_{\alpha}$ being real, and obtain
\begin{equation}
 {\cal R} = \frac{\hbar}{2}\left[\frac{1+|\beta|^2-2|\beta|\sin\phi_{\alpha}'}
 {1+|\beta|^2+2|\beta|\sin\phi_{\alpha}'}\right]^{1/2}\,,
\end{equation}
in which we have introduced
\begin{equation}\label{eq:phi_alpha}
\phi_{\alpha}' \equiv \phi_{\alpha}+\arctan[{\rm Re}
(\beta)/{\rm Im}(\beta)]\,.
\end{equation}
Due to the one-to-one mapping between
$\theta$ and $\phi'_{\alpha}$, minimizing $\cal R$ over $\theta$
is therefore equivalent to that over $\phi'_{\alpha}$.
The minimum of $\cal R$
is achieved when $\phi'_{\alpha}=\pi/2$ and
\begin{equation}
 {\cal R}_{\rm min}=\frac{\hbar}{2}\left|\frac{1-|\beta|}
 {1+|\beta|}\right|\,.
\end{equation}
It is always smaller than $\hbar/2$, i.e.,
\begin{equation}
  {\cal R}_{\rm min}\le \frac{\hbar}{2}\,,
\end{equation}
and reaches the equal sign when either
\begin{equation}
|\beta|=0\quad{\rm or}\quad |\beta|\rightarrow \infty\,.
\end{equation}
From the definition of $\beta$
Eq.\,\eqref{eq:beta},
this corresponds to either
${\mathscr F}(-\omega)=0$ or ${\mathscr F}(\omega)=0$, which is equivalent to
\begin{equation}
S_{FF}(-\omega)=0\quad {\rm or}\quad S_{FF}(\omega)=0\,,
\end{equation}
 according to Eq.\,\eqref{eq:spec_FF}.
With the same argument as the one
presented in the previous section,
the above conclusion is not conditional on whether the detector is
in the vacuum state or in the general, stationary, pure Gaussian state.

Q.E.D.


%


\begin{thebibliography}{38}%
\makeatletter
\providecommand \@ifxundefined [1]{%
 \@ifx{#1\undefined}
}%
\providecommand \@ifnum [1]{%
 \ifnum #1\expandafter \@firstoftwo
 \else \expandafter \@secondoftwo
 \fi
}%
\providecommand \@ifx [1]{%
 \ifx #1\expandafter \@firstoftwo
 \else \expandafter \@secondoftwo
 \fi
}%
\providecommand \natexlab [1]{#1}%
\providecommand \enquote  [1]{``#1''}%
\providecommand \bibnamefont  [1]{#1}%
\providecommand \bibfnamefont [1]{#1}%
\providecommand \citenamefont [1]{#1}%
\providecommand \href@noop [0]{\@secondoftwo}%
\providecommand \href [0]{\begingroup \@sanitize@url \@href}%
\providecommand \@href[1]{\@@startlink{#1}\@@href}%
\providecommand \@@href[1]{\endgroup#1\@@endlink}%
\providecommand \@sanitize@url [0]{\catcode `\\12\catcode `\$12\catcode
  `\&12\catcode `\#12\catcode `\^12\catcode `\_12\catcode `\%12\relax}%
\providecommand \@@startlink[1]{}%
\providecommand \@@endlink[0]{}%
\providecommand \url  [0]{\begingroup\@sanitize@url \@url }%
\providecommand \@url [1]{\endgroup\@href {#1}{\urlprefix }}%
\providecommand \urlprefix  [0]{URL }%
\providecommand \Eprint [0]{\href }%
\providecommand \doibase [0]{http://dx.doi.org/}%
\providecommand \selectlanguage [0]{\@gobble}%
\providecommand \bibinfo  [0]{\@secondoftwo}%
\providecommand \bibfield  [0]{\@secondoftwo}%
\providecommand \translation [1]{[#1]}%
\providecommand \BibitemOpen [0]{}%
\providecommand \bibitemStop [0]{}%
\providecommand \bibitemNoStop [0]{.\EOS\space}%
\providecommand \EOS [0]{\spacefactor3000\relax}%
\providecommand \BibitemShut  [1]{\csname bibitem#1\endcsname}%
\let\auto@bib@innerbib\@empty
\bibitem [{\citenamefont {Giovannetti}\ \emph {et~al.}(2011)\citenamefont
  {Giovannetti}, \citenamefont {Lloyd},\ and\ \citenamefont
  {Maccone}}]{Giovannetti2011}%
  \BibitemOpen
  \bibfield  {author} {\bibinfo {author} {\bibfnamefont {V.}~\bibnamefont
  {Giovannetti}}, \bibinfo {author} {\bibfnamefont {S.}~\bibnamefont {Lloyd}},
  \ and\ \bibinfo {author} {\bibfnamefont {L.}~\bibnamefont {Maccone}},\ }\href
  {\doibase 10.1038/nphoton.2011.35} {\bibfield  {journal} {\bibinfo  {journal}
  {Nature Photonics}\ }\textbf {\bibinfo {volume} {5}},\ \bibinfo {pages} {222}
  (\bibinfo {year} {2011})}\BibitemShut {NoStop}%
\bibitem [{\citenamefont {Helstrom}(1967)}]{Helstrom1967}%
  \BibitemOpen
  \bibfield  {author} {\bibinfo {author} {\bibfnamefont {C.}~\bibnamefont
  {Helstrom}},\ }\href {\doibase 10.1016/0375-9601(67)90366-0} {\bibfield
  {journal} {\bibinfo  {journal} {Phys. Lett. A}\ }\textbf {\bibinfo {volume}
  {25}},\ \bibinfo {pages} {101} (\bibinfo {year} {1967})}\BibitemShut
  {NoStop}%
\bibitem [{\citenamefont {Holevo}(2011)}]{Holevo2011}%
  \BibitemOpen
  \bibfield  {author} {\bibinfo {author} {\bibfnamefont {A.}~\bibnamefont
  {Holevo}},\ }\href {\doibase 10.1007/978-88-7642-378-9} {\emph {\bibinfo
  {title} {{Probabilistic and Statistical Aspects of quantum theory}}}},\
  \bibinfo {edition} {2nd}\ ed.\ (\bibinfo  {publisher} {Scuola Normale
  Superiore},\ \bibinfo {year} {2011})\BibitemShut {NoStop}%
\bibitem [{\citenamefont {Braunstein}\ and\ \citenamefont
  {Caves}(1994)}]{Braunstein1994}%
  \BibitemOpen
  \bibfield  {author} {\bibinfo {author} {\bibfnamefont {S.~L.}\ \bibnamefont
  {Braunstein}}\ and\ \bibinfo {author} {\bibfnamefont {C.~M.}\ \bibnamefont
  {Caves}},\ }\href {\doibase 10.1103/PhysRevLett.72.3439} {\bibfield
  {journal} {\bibinfo  {journal} {Phys. Rev. Lett.}\ }\textbf {\bibinfo
  {volume} {72}},\ \bibinfo {pages} {3439} (\bibinfo {year}
  {1994})}\BibitemShut {NoStop}%
\bibitem [{\citenamefont {Braunstein}\ \emph {et~al.}(1996)\citenamefont
  {Braunstein}, \citenamefont {Caves},\ and\ \citenamefont
  {Milburn}}]{Braunstein1996}%
  \BibitemOpen
  \bibfield  {author} {\bibinfo {author} {\bibfnamefont {S.~L.}\ \bibnamefont
  {Braunstein}}, \bibinfo {author} {\bibfnamefont {C.~M.}\ \bibnamefont
  {Caves}}, \ and\ \bibinfo {author} {\bibfnamefont {G.~J.}\ \bibnamefont
  {Milburn}},\ }\href {\doibase 10.1006/aphy.1996.0040} {\bibfield  {journal}
  {\bibinfo  {journal} {Annals of Physics}\ }\textbf {\bibinfo {volume}
  {247}},\ \bibinfo {pages} {135} (\bibinfo {year} {1996})}\BibitemShut
  {NoStop}%
\bibitem [{\citenamefont {Wiseman}\ and\ \citenamefont
  {Milburn}(2010)}]{Wiseman2010}%
  \BibitemOpen
  \bibfield  {author} {\bibinfo {author} {\bibfnamefont {H.~M.}\ \bibnamefont
  {Wiseman}}\ and\ \bibinfo {author} {\bibfnamefont {G.~J.}\ \bibnamefont
  {Milburn}},\ }\href {\doibase 10.1017/CBO9780511813948} {\emph {\bibinfo
  {title} {{Quantum Measurement and Control}}}}\ (\bibinfo  {publisher}
  {Cambridge University Press},\ \bibinfo {year} {2010})\BibitemShut {NoStop}%
\bibitem [{Note1()}]{Note1}%
  \BibitemOpen
  \bibinfo {note} {Due to the central limit theorem. In this case, an extra
  factor of $1/N$ ($N$ the sample size) shall be included in the bound
  above.}\BibitemShut {Stop}%
\bibitem [{\citenamefont {Szczykulska}\ \emph {et~al.}(2016)\citenamefont
  {Szczykulska}, \citenamefont {Baumgratz},\ and\ \citenamefont
  {Datta}}]{Szczykulska2016b}%
  \BibitemOpen
  \bibfield  {author} {\bibinfo {author} {\bibfnamefont {M.}~\bibnamefont
  {Szczykulska}}, \bibinfo {author} {\bibfnamefont {T.}~\bibnamefont
  {Baumgratz}}, \ and\ \bibinfo {author} {\bibfnamefont {A.}~\bibnamefont
  {Datta}},\ }\href {\doibase 10.1080/23746149.2016.1230476} {\bibfield
  {journal} {\bibinfo  {journal} {Advances in Physics: X}\ }\textbf {\bibinfo
  {volume} {1}},\ \bibinfo {pages} {621} (\bibinfo {year} {2016})}\BibitemShut
  {NoStop}%
\bibitem [{\citenamefont {Tsang}\ \emph {et~al.}(2011)\citenamefont {Tsang},
  \citenamefont {Wiseman},\ and\ \citenamefont {Caves}}]{Tsang2011}%
  \BibitemOpen
  \bibfield  {author} {\bibinfo {author} {\bibfnamefont {M.}~\bibnamefont
  {Tsang}}, \bibinfo {author} {\bibfnamefont {H.~M.}\ \bibnamefont {Wiseman}},
  \ and\ \bibinfo {author} {\bibfnamefont {C.~M.}\ \bibnamefont {Caves}},\
  }\href {\doibase 10.1103/PhysRevLett.106.090401} {\bibfield  {journal}
  {\bibinfo  {journal} {Phys. Rev. Lett.}\ }\textbf {\bibinfo {volume} {106}},\
  \bibinfo {pages} {090401} (\bibinfo {year} {2011})}\BibitemShut {NoStop}%
\bibitem [{\citenamefont {{C. M. Caves}}(1981)}]{Caves81}%
  \BibitemOpen
  \bibfield  {author} {\bibinfo {author} {\bibnamefont {{C. M. Caves}}},\
  }\href {http://journals.aps.org/prd/abstract/10.1103/PhysRevD.23.1693}
  {\bibfield  {journal} {\bibinfo  {journal} {Phys. Rev. D}\ }\textbf {\bibinfo
  {volume} {23}},\ \bibinfo {pages} {1693} (\bibinfo {year}
  {1981})}\BibitemShut {NoStop}%
\bibitem [{\citenamefont {Adhikari}(2014)}]{Adhikari2014}%
  \BibitemOpen
  \bibfield  {author} {\bibinfo {author} {\bibfnamefont {R.~X.}\ \bibnamefont
  {Adhikari}},\ }\href
  {http://journals.aps.org/rmp/abstract/10.1103/RevModPhys.86.121} {\bibfield
  {journal} {\bibinfo  {journal} {Rev. Mod. Phys.}\ }\textbf {\bibinfo {volume}
  {86}},\ \bibinfo {pages} {121} (\bibinfo {year} {2014})}\BibitemShut
  {NoStop}%
\bibitem [{\citenamefont {Braginsky}\ \emph {et~al.}(2000)\citenamefont
  {Braginsky}, \citenamefont {Gorodetsky}, \citenamefont {Khalili},\ and\
  \citenamefont {Thorne}}]{Braginsky2000EQL}%
  \BibitemOpen
  \bibfield  {author} {\bibinfo {author} {\bibfnamefont {V.~B.}\ \bibnamefont
  {Braginsky}}, \bibinfo {author} {\bibfnamefont {M.~L.}\ \bibnamefont
  {Gorodetsky}}, \bibinfo {author} {\bibfnamefont {F.~Y.}\ \bibnamefont
  {Khalili}}, \ and\ \bibinfo {author} {\bibfnamefont {K.~S.}\ \bibnamefont
  {Thorne}},\ }\href
  {http://scitation.aip.org/content/aip/proceeding/aipcp/10.1063/1.1291855}
  {\bibfield  {journal} {\bibinfo  {journal} {AIP Conf. Proc.}\ }\textbf
  {\bibinfo {volume} {523}},\ \bibinfo {pages} {180} (\bibinfo {year}
  {2000})}\BibitemShut {NoStop}%
\bibitem [{\citenamefont {Kubo}(1966)}]{Kubo1966}%
  \BibitemOpen
  \bibfield  {author} {\bibinfo {author} {\bibfnamefont {R.}~\bibnamefont
  {Kubo}},\ }\href {http://stacks.iop.org/0034-4885/29/i=1/a=306} {\bibfield
  {journal} {\bibinfo  {journal} {Reports on Progress in Physics}\ }\textbf
  {\bibinfo {volume} {29}},\ \bibinfo {pages} {255} (\bibinfo {year}
  {1966})}\BibitemShut {NoStop}%
\bibitem [{\citenamefont {Averin}(2003)}]{Averin2003}%
  \BibitemOpen
  \bibfield  {author} {\bibinfo {author} {\bibfnamefont {D.~V.}\ \bibnamefont
  {Averin}},\ }\href {http://arxiv.org/abs/cond-mat/0301524} {\bibfield
  {journal} {\bibinfo  {journal} {arXiv:cond-mat/0301524}\ } (\bibinfo {year}
  {2003})}\BibitemShut {NoStop}%
\bibitem [{\citenamefont {Clerk}\ \emph {et~al.}(2003)\citenamefont {Clerk},
  \citenamefont {Girvin},\ and\ \citenamefont {Stone}}]{Clerk2003}%
  \BibitemOpen
  \bibfield  {author} {\bibinfo {author} {\bibfnamefont {A.~A.}\ \bibnamefont
  {Clerk}}, \bibinfo {author} {\bibfnamefont {S.~M.}\ \bibnamefont {Girvin}}, \
  and\ \bibinfo {author} {\bibfnamefont {A.~D.}\ \bibnamefont {Stone}},\ }\href
  {\doibase 10.1103/PhysRevB.67.165324} {\bibfield  {journal} {\bibinfo
  {journal} {Phys. Rev. B}\ }\textbf {\bibinfo {volume} {67}},\ \bibinfo
  {pages} {165324} (\bibinfo {year} {2003})}\BibitemShut {NoStop}%
\bibitem [{\citenamefont {Braginsky}\ and\ \citenamefont
  {Khalilli}(1992)}]{Braginsky92}%
  \BibitemOpen
  \bibfield  {author} {\bibinfo {author} {\bibfnamefont {V.~B.}\ \bibnamefont
  {Braginsky}}\ and\ \bibinfo {author} {\bibfnamefont {F.}~\bibnamefont
  {Khalilli}},\ }\href
  {http://www.cambridge.org/us/academic/subjects/physics/quantum-physics-quant%
um-information-and-quantum-computation/quantum-measurement} {\emph {\bibinfo
  {title} {{Quantum Measurement}}}}\ (\bibinfo  {publisher} {Cambridge
  University Press},\ \bibinfo {year} {1992})\BibitemShut {NoStop}%
\bibitem [{\citenamefont {Clerk}\ \emph {et~al.}(2010)\citenamefont {Clerk},
  \citenamefont {Devoret}, \citenamefont {Girvin}, \citenamefont {Marquardt},\
  and\ \citenamefont {Schoelkopf}}]{Clerk2008}%
  \BibitemOpen
  \bibfield  {author} {\bibinfo {author} {\bibfnamefont {A.~A.}\ \bibnamefont
  {Clerk}}, \bibinfo {author} {\bibfnamefont {M.~H.}\ \bibnamefont {Devoret}},
  \bibinfo {author} {\bibfnamefont {S.~M.}\ \bibnamefont {Girvin}}, \bibinfo
  {author} {\bibfnamefont {F.}~\bibnamefont {Marquardt}}, \ and\ \bibinfo
  {author} {\bibfnamefont {R.~J.}\ \bibnamefont {Schoelkopf}},\ }\href
  {\doibase 10.1103/RevModPhys.82.1155} {\bibfield  {journal} {\bibinfo
  {journal} {Rev. Mod. Phys.}\ }\textbf {\bibinfo {volume} {82}},\ \bibinfo
  {pages} {1155} (\bibinfo {year} {2010})}\BibitemShut {NoStop}%
\bibitem [{\citenamefont {Miao}(2017)}]{Miao2017}%
  \BibitemOpen
  \bibfield  {author} {\bibinfo {author} {\bibfnamefont {H.}~\bibnamefont
  {Miao}},\ }\href {\doibase 10.1103/PhysRevA.95.012103} {\bibfield  {journal}
  {\bibinfo  {journal} {Phys. Rev. A}\ }\textbf {\bibinfo {volume} {95}},\
  \bibinfo {pages} {012103} (\bibinfo {year} {2017})}\BibitemShut {NoStop}%
\bibitem [{\citenamefont {Blow}\ \emph {et~al.}(1990)\citenamefont {Blow},
  \citenamefont {Loudon}, \citenamefont {Phoenix},\ and\ \citenamefont
  {Shepherd}}]{Blow1990}%
  \BibitemOpen
  \bibfield  {author} {\bibinfo {author} {\bibfnamefont {K.~J.}\ \bibnamefont
  {Blow}}, \bibinfo {author} {\bibfnamefont {R.}~\bibnamefont {Loudon}},
  \bibinfo {author} {\bibfnamefont {S.~J.~D.}\ \bibnamefont {Phoenix}}, \ and\
  \bibinfo {author} {\bibfnamefont {T.~J.}\ \bibnamefont {Shepherd}},\ }\href
  {\doibase 10.1103/PhysRevA.42.4102} {\bibfield  {journal} {\bibinfo
  {journal} {Phys. Rev. A}\ }\textbf {\bibinfo {volume} {42}},\ \bibinfo
  {pages} {4102} (\bibinfo {year} {1990})}\BibitemShut {NoStop}%
\bibitem [{Note2()}]{Note2}%
  \BibitemOpen
  \bibinfo {note} {In addition to Refs.\protect \tmspace +\thinmuskip
  {.1667em}\cite {Kubo1966, Braginsky92, BuCh2002,Clerk2008, Blow1990}
  mentioned in the main text, the supplemental material also includes
  Ref.\protect \tmspace +\thinmuskip {.1667em}\cite
  {Gardiner1985}.}\BibitemShut {Stop}%
\bibitem [{\citenamefont {Braginsky}\ and\ \citenamefont
  {Khalili}(1996)}]{Braginsky1996}%
  \BibitemOpen
  \bibfield  {author} {\bibinfo {author} {\bibfnamefont {V.~B.}\ \bibnamefont
  {Braginsky}}\ and\ \bibinfo {author} {\bibfnamefont {F.~Y.}\ \bibnamefont
  {Khalili}},\ }\href {\doibase 10.1103/RevModPhys.68.1} {\bibfield  {journal}
  {\bibinfo  {journal} {Reviews of Modern Physics}\ }\textbf {\bibinfo {volume}
  {68}},\ \bibinfo {pages} {1} (\bibinfo {year} {1996})}\BibitemShut {NoStop}%
\bibitem [{\citenamefont {Tsang}\ and\ \citenamefont {Caves}(2010)}]{Tsang10}%
  \BibitemOpen
  \bibfield  {author} {\bibinfo {author} {\bibfnamefont {M.}~\bibnamefont
  {Tsang}}\ and\ \bibinfo {author} {\bibfnamefont {C.~M.}\ \bibnamefont
  {Caves}},\ }\href {\doibase 10.1103/PhysRevLett.105.123601} {\bibfield
  {journal} {\bibinfo  {journal} {Phys. Rev. Lett.}\ }\textbf {\bibinfo
  {volume} {105}},\ \bibinfo {pages} {123601} (\bibinfo {year}
  {2010})}\BibitemShut {NoStop}%
\bibitem [{\citenamefont {{H. J. Kimble}}\ \emph {et~al.}(2001)\citenamefont
  {{H. J. Kimble}}, \citenamefont {{Y. Levin}}, \citenamefont {{A. B. Matsko}},
  \citenamefont {{K. S. Thorne}},\ and\ \citenamefont {{S. P.
  Vyatchanin}}}]{KLMTV2001}%
  \BibitemOpen
  \bibfield  {author} {\bibinfo {author} {\bibnamefont {{H. J. Kimble}}},
  \bibinfo {author} {\bibnamefont {{Y. Levin}}}, \bibinfo {author}
  {\bibnamefont {{A. B. Matsko}}}, \bibinfo {author} {\bibnamefont {{K. S.
  Thorne}}}, \ and\ \bibinfo {author} {\bibnamefont {{S. P. Vyatchanin}}},\
  }\href {http://journals.aps.org/prd/abstract/10.1103/PhysRevD.65.022002}
  {\bibfield  {journal} {\bibinfo  {journal} {Phys. Rev. D}\ }\textbf {\bibinfo
  {volume} {65}},\ \bibinfo {pages} {022002} (\bibinfo {year}
  {2001})}\BibitemShut {NoStop}%
\bibitem [{\citenamefont {{LIGO Scientific
  Collaboration}}(2015)}]{AdvancedLIGO15}%
  \BibitemOpen
  \bibfield  {author} {\bibinfo {author} {\bibnamefont {{LIGO Scientific
  Collaboration}}},\ }\href {http://stacks.iop.org/0264-9381/32/i=7/a=074001}
  {\bibfield  {journal} {\bibinfo  {journal} {Classical and Quantum Gravity}\
  }\textbf {\bibinfo {volume} {32}},\ \bibinfo {pages} {74001} (\bibinfo {year}
  {2015})}\BibitemShut {NoStop}%
\bibitem [{\citenamefont {Cooperstock}\ and\ \citenamefont
  {Faraoni}(1993)}]{Cooperstock1993}%
  \BibitemOpen
  \bibfield  {author} {\bibinfo {author} {\bibfnamefont {F.~I.}\ \bibnamefont
  {Cooperstock}}\ and\ \bibinfo {author} {\bibfnamefont {V.}~\bibnamefont
  {Faraoni}},\ }\href {\doibase 10.1088/0264-9381/10/6/016} {\bibfield
  {journal} {\bibinfo  {journal} {Classical and Quantum Gravity}\ }\textbf
  {\bibinfo {volume} {10}},\ \bibinfo {pages} {1189} (\bibinfo {year}
  {1993})}\BibitemShut {NoStop}%
\bibitem [{\citenamefont {Tarabrin}\ and\ \citenamefont
  {Seleznyov}(2008)}]{Tarabrin08}%
  \BibitemOpen
  \bibfield  {author} {\bibinfo {author} {\bibfnamefont {S.~P.}\ \bibnamefont
  {Tarabrin}}\ and\ \bibinfo {author} {\bibfnamefont {A.~A.}\ \bibnamefont
  {Seleznyov}},\ }\href {\doibase 10.1103/PhysRevD.78.062001} {\bibfield
  {journal} {\bibinfo  {journal} {Phys. Rev. D}\ }\textbf {\bibinfo {volume}
  {78}},\ \bibinfo {pages} {062001} (\bibinfo {year} {2008})}\BibitemShut
  {NoStop}%
\bibitem [{\citenamefont {Buonanno}\ and\ \citenamefont
  {Chen}(2003)}]{scaling_law}%
  \BibitemOpen
  \bibfield  {author} {\bibinfo {author} {\bibfnamefont {A.}~\bibnamefont
  {Buonanno}}\ and\ \bibinfo {author} {\bibfnamefont {Y.}~\bibnamefont
  {Chen}},\ }\href
  {http://journals.aps.org/prd/abstract/10.1103/PhysRevD.67.062002} {\bibfield
  {journal} {\bibinfo  {journal} {Phys. Rev. D}\ }\textbf {\bibinfo {volume}
  {67}},\ \bibinfo {pages} {062002} (\bibinfo {year} {2003})}\BibitemShut
  {NoStop}%
\bibitem [{\citenamefont {Chen}(2013)}]{Yanbei:review}%
  \BibitemOpen
  \bibfield  {author} {\bibinfo {author} {\bibfnamefont {Y.}~\bibnamefont
  {Chen}},\ }\href
  {http://iopscience.iop.org/article/10.1088/0953-4075/46/10/104001/meta}
  {\bibfield  {journal} {\bibinfo  {journal} {Journal of Physics B: Atomic,
  Molecular and Optical Physics}\ }\textbf {\bibinfo {volume} {46}},\ \bibinfo
  {pages} {104001} (\bibinfo {year} {2013})}\BibitemShut {NoStop}%
\bibitem [{\citenamefont {Aspelmeyer}\ \emph {et~al.}(2014)\citenamefont
  {Aspelmeyer}, \citenamefont {Kippenberg},\ and\ \citenamefont
  {Marquardt}}]{Aspelmeyer2014}%
  \BibitemOpen
  \bibfield  {author} {\bibinfo {author} {\bibfnamefont {M.}~\bibnamefont
  {Aspelmeyer}}, \bibinfo {author} {\bibfnamefont {T.~J.}\ \bibnamefont
  {Kippenberg}}, \ and\ \bibinfo {author} {\bibfnamefont {F.}~\bibnamefont
  {Marquardt}},\ }\href {\doibase 10.1103/RevModPhys.86.1391} {\bibfield
  {journal} {\bibinfo  {journal} {Rev. Mod. Phys.}\ }\textbf {\bibinfo {volume}
  {86}},\ \bibinfo {pages} {1391} (\bibinfo {year} {2014})}\BibitemShut
  {NoStop}%
\bibitem [{\citenamefont {Harms}\ \emph {et~al.}(2003)\citenamefont {Harms},
  \citenamefont {Chen}, \citenamefont {Chelkowski}, \citenamefont {Franzen},
  \citenamefont {Vahlbruch}, \citenamefont {Danzmann},\ and\ \citenamefont
  {Schnabel}}]{Harms2003a}%
  \BibitemOpen
  \bibfield  {author} {\bibinfo {author} {\bibfnamefont {J.}~\bibnamefont
  {Harms}}, \bibinfo {author} {\bibfnamefont {Y.}~\bibnamefont {Chen}},
  \bibinfo {author} {\bibfnamefont {S.}~\bibnamefont {Chelkowski}}, \bibinfo
  {author} {\bibfnamefont {A.}~\bibnamefont {Franzen}}, \bibinfo {author}
  {\bibfnamefont {H.}~\bibnamefont {Vahlbruch}}, \bibinfo {author}
  {\bibfnamefont {K.}~\bibnamefont {Danzmann}}, \ and\ \bibinfo {author}
  {\bibfnamefont {R.}~\bibnamefont {Schnabel}},\ }\href {\doibase
  10.1103/PhysRevD.68.042001} {\bibfield  {journal} {\bibinfo  {journal} {Phys.
  Rev. D}\ }\textbf {\bibinfo {volume} {68}},\ \bibinfo {pages} {042001}
  (\bibinfo {year} {2003})}\BibitemShut {NoStop}%
\bibitem [{\citenamefont {Corbitt}\ \emph {et~al.}(2006)\citenamefont
  {Corbitt}, \citenamefont {Chen}, \citenamefont {Khalili}, \citenamefont
  {Ottaway}, \citenamefont {Vyatchanin}, \citenamefont {Whitcomb},\ and\
  \citenamefont {Mavalvala}}]{Corbitt2006}%
  \BibitemOpen
  \bibfield  {author} {\bibinfo {author} {\bibfnamefont {T.}~\bibnamefont
  {Corbitt}}, \bibinfo {author} {\bibfnamefont {Y.}~\bibnamefont {Chen}},
  \bibinfo {author} {\bibfnamefont {F.}~\bibnamefont {Khalili}}, \bibinfo
  {author} {\bibfnamefont {D.}~\bibnamefont {Ottaway}}, \bibinfo {author}
  {\bibfnamefont {S.}~\bibnamefont {Vyatchanin}}, \bibinfo {author}
  {\bibfnamefont {S.}~\bibnamefont {Whitcomb}}, \ and\ \bibinfo {author}
  {\bibfnamefont {N.}~\bibnamefont {Mavalvala}},\ }\href
  {http://link.aps.org/doi/10.1103/PhysRevA.73.023801} {\bibfield  {journal}
  {\bibinfo  {journal} {Phys. Rev. A}\ }\textbf {\bibinfo {volume} {73}},\
  \bibinfo {pages} {023801} (\bibinfo {year} {2006})}\BibitemShut {NoStop}%
\bibitem [{\citenamefont {Brooks}\ \emph {et~al.}(2012)\citenamefont {Brooks},
  \citenamefont {Botter}, \citenamefont {Schreppler}, \citenamefont {Purdy},
  \citenamefont {Brahms},\ and\ \citenamefont {Stamper-Kurn}}]{Brooks2012}%
  \BibitemOpen
  \bibfield  {author} {\bibinfo {author} {\bibfnamefont {D.~W.~C.}\
  \bibnamefont {Brooks}}, \bibinfo {author} {\bibfnamefont {T.}~\bibnamefont
  {Botter}}, \bibinfo {author} {\bibfnamefont {S.}~\bibnamefont {Schreppler}},
  \bibinfo {author} {\bibfnamefont {T.~P.}\ \bibnamefont {Purdy}}, \bibinfo
  {author} {\bibfnamefont {N.}~\bibnamefont {Brahms}}, \ and\ \bibinfo {author}
  {\bibfnamefont {D.~M.}\ \bibnamefont {Stamper-Kurn}},\ }\href
  {http://dx.doi.org/10.1038/nature11325} {\bibfield  {journal} {\bibinfo
  {journal} {Nature}\ }\textbf {\bibinfo {volume} {488}},\ \bibinfo {pages}
  {476} (\bibinfo {year} {2012})}\BibitemShut {NoStop}%
\bibitem [{\citenamefont {Purdy}\ \emph {et~al.}(2013)\citenamefont {Purdy},
  \citenamefont {Yu}, \citenamefont {Peterson}, \citenamefont {Kampel},\ and\
  \citenamefont {Regal}}]{Purdy2013}%
  \BibitemOpen
  \bibfield  {author} {\bibinfo {author} {\bibfnamefont {T.~P.}\ \bibnamefont
  {Purdy}}, \bibinfo {author} {\bibfnamefont {P.~L.}\ \bibnamefont {Yu}},
  \bibinfo {author} {\bibfnamefont {R.~W.}\ \bibnamefont {Peterson}}, \bibinfo
  {author} {\bibfnamefont {N.~S.}\ \bibnamefont {Kampel}}, \ and\ \bibinfo
  {author} {\bibfnamefont {C.~A.}\ \bibnamefont {Regal}},\ }\href
  {http://journals.aps.org/prx/abstract/10.1103/PhysRevX.3.031012} {\bibfield
  {journal} {\bibinfo  {journal} {Phys. Rev. X}\ }\textbf {\bibinfo {volume}
  {3}},\ \bibinfo {pages} {031012} (\bibinfo {year} {2013})}\BibitemShut
  {NoStop}%
\bibitem [{\citenamefont {Safavi-Naeini}\ \emph {et~al.}(2013)\citenamefont
  {Safavi-Naeini}, \citenamefont {Gr{\"{o}}blacher}, \citenamefont {Hill},
  \citenamefont {Chan}, \citenamefont {Aspelmeyer},\ and\ \citenamefont
  {Painter}}]{Safavi-Naeini2013a}%
  \BibitemOpen
  \bibfield  {author} {\bibinfo {author} {\bibfnamefont {A.~H.}\ \bibnamefont
  {Safavi-Naeini}}, \bibinfo {author} {\bibfnamefont {S.}~\bibnamefont
  {Gr{\"{o}}blacher}}, \bibinfo {author} {\bibfnamefont {J.~T.}\ \bibnamefont
  {Hill}}, \bibinfo {author} {\bibfnamefont {J.}~\bibnamefont {Chan}}, \bibinfo
  {author} {\bibfnamefont {M.}~\bibnamefont {Aspelmeyer}}, \ and\ \bibinfo
  {author} {\bibfnamefont {O.}~\bibnamefont {Painter}},\ }\href {\doibase
  10.1038/nature12307} {\bibfield  {journal} {\bibinfo  {journal} {Nature}\
  }\textbf {\bibinfo {volume} {500}},\ \bibinfo {pages} {185} (\bibinfo {year}
  {2013})}\BibitemShut {NoStop}%
\bibitem [{\citenamefont {Peano}\ \emph {et~al.}(2015)\citenamefont {Peano},
  \citenamefont {Schwefel}, \citenamefont {Marquardt},\ and\ \citenamefont
  {Marquardt}}]{Peano2015}%
  \BibitemOpen
  \bibfield  {author} {\bibinfo {author} {\bibfnamefont {V.}~\bibnamefont
  {Peano}}, \bibinfo {author} {\bibfnamefont {H.~G.~L.}\ \bibnamefont
  {Schwefel}}, \bibinfo {author} {\bibfnamefont {C.}~\bibnamefont {Marquardt}},
  \ and\ \bibinfo {author} {\bibfnamefont {F.}~\bibnamefont {Marquardt}},\
  }\href {http://journals.aps.org/prl/abstract/10.1103/PhysRevLett.115.243603}
  {\bibfield  {journal} {\bibinfo  {journal} {Phys. Rev. Lett.}\ }\textbf
  {\bibinfo {volume} {115}},\ \bibinfo {pages} {243603} (\bibinfo {year}
  {2015})}\BibitemShut {NoStop}%
\bibitem [{\citenamefont {Korobko}\ \emph {et~al.}(2017)\citenamefont
  {Korobko}, \citenamefont {Kleybolte}, \citenamefont {Ast}, \citenamefont
  {Miao}, \citenamefont {Chen},\ and\ \citenamefont {Schnabel}}]{Korobko2017}%
  \BibitemOpen
  \bibfield  {author} {\bibinfo {author} {\bibfnamefont {M.}~\bibnamefont
  {Korobko}}, \bibinfo {author} {\bibfnamefont {L.}~\bibnamefont {Kleybolte}},
  \bibinfo {author} {\bibfnamefont {S.}~\bibnamefont {Ast}}, \bibinfo {author}
  {\bibfnamefont {H.}~\bibnamefont {Miao}}, \bibinfo {author} {\bibfnamefont
  {Y.}~\bibnamefont {Chen}}, \ and\ \bibinfo {author} {\bibfnamefont
  {R.}~\bibnamefont {Schnabel}},\ }\href {\doibase
  10.1103/PhysRevLett.118.143601} {\bibfield  {journal} {\bibinfo  {journal}
  {Phys. Rev. Lett.}\ }\textbf {\bibinfo {volume} {118}},\ \bibinfo {pages}
  {143601} (\bibinfo {year} {2017})}\BibitemShut {NoStop}%
\bibitem [{\citenamefont {{A. Buonanno}}\ and\ \citenamefont {{Y.
  Chen}}(2002)}]{BuCh2002}%
  \BibitemOpen
  \bibfield  {author} {\bibinfo {author} {\bibnamefont {{A. Buonanno}}}\ and\
  \bibinfo {author} {\bibnamefont {{Y. Chen}}},\ }\href
  {http://journals.aps.org/prd/abstract/10.1103/PhysRevD.65.042001} {\bibfield
  {journal} {\bibinfo  {journal} {Phys. Rev. D}\ }\textbf {\bibinfo {volume}
  {65}},\ \bibinfo {pages} {042001} (\bibinfo {year} {2002})}\BibitemShut
  {NoStop}%
\bibitem [{\citenamefont {Gardiner}\ and\ \citenamefont
  {Collett}(1985)}]{Gardiner1985}%
  \BibitemOpen
  \bibfield  {author} {\bibinfo {author} {\bibfnamefont {C.~W.}\ \bibnamefont
  {Gardiner}}\ and\ \bibinfo {author} {\bibfnamefont {M.~J.}\ \bibnamefont
  {Collett}},\ }\href {\doibase 10.1103/PhysRevA.31.3761} {\bibfield  {journal}
  {\bibinfo  {journal} {Phys. Rev. A}\ }\textbf {\bibinfo {volume} {31}},\
  \bibinfo {pages} {3761} (\bibinfo {year} {1985})}\BibitemShut {NoStop}%
\end{thebibliography}

\begin{thebibliography}{6}%
\makeatletter
\providecommand \@ifxundefined [1]{%
 \@ifx{#1\undefined}
}%
\providecommand \@ifnum [1]{%
 \ifnum #1\expandafter \@firstoftwo
 \else \expandafter \@secondoftwo
 \fi
}%
\providecommand \@ifx [1]{%
 \ifx #1\expandafter \@firstoftwo
 \else \expandafter \@secondoftwo
 \fi
}%
\providecommand \natexlab [1]{#1}%
\providecommand \enquote  [1]{``#1''}%
\providecommand \bibnamefont  [1]{#1}%
\providecommand \bibfnamefont [1]{#1}%
\providecommand \citenamefont [1]{#1}%
\providecommand \href@noop [0]{\@secondoftwo}%
\providecommand \href [0]{\begingroup \@sanitize@url \@href}%
\providecommand \@href[1]{\@@startlink{#1}\@@href}%
\providecommand \@@href[1]{\endgroup#1\@@endlink}%
\providecommand \@sanitize@url [0]{\catcode `\\12\catcode `\$12\catcode
  `\&12\catcode `\#12\catcode `\^12\catcode `\_12\catcode `\%12\relax}%
\providecommand \@@startlink[1]{}%
\providecommand \@@endlink[0]{}%
\providecommand \url  [0]{\begingroup\@sanitize@url \@url }%
\providecommand \@url [1]{\endgroup\@href {#1}{\urlprefix }}%
\providecommand \urlprefix  [0]{URL }%
\providecommand \Eprint [0]{\href }%
\providecommand \doibase [0]{http://dx.doi.org/}%
\providecommand \selectlanguage [0]{\@gobble}%
\providecommand \bibinfo  [0]{\@secondoftwo}%
\providecommand \bibfield  [0]{\@secondoftwo}%
\providecommand \translation [1]{[#1]}%
\providecommand \BibitemOpen [0]{}%
\providecommand \bibitemStop [0]{}%
\providecommand \bibitemNoStop [0]{.\EOS\space}%
\providecommand \EOS [0]{\spacefactor3000\relax}%
\providecommand \BibitemShut  [1]{\csname bibitem#1\endcsname}%
\let\auto@bib@innerbib\@empty
\bibitem [{\citenamefont {Kubo}(1966)}]{Kubo1966s}%
  \BibitemOpen
  \bibfield  {author} {\bibinfo {author} {\bibfnamefont {R.}~\bibnamefont
  {Kubo}},\ }\href {http://stacks.iop.org/0034-4885/29/i=1/a=306} {\bibfield
  {journal} {\bibinfo  {journal} {Reports on Progress in Physics}\ }\textbf
  {\bibinfo {volume} {29}},\ \bibinfo {pages} {255} (\bibinfo {year}
  {1966})}\BibitemShut {NoStop}%
\bibitem [{\citenamefont {Braginsky}\ and\ \citenamefont
  {Khalilli}(1992)}]{Braginsky92s}%
  \BibitemOpen
  \bibfield  {author} {\bibinfo {author} {\bibfnamefont {V.~B.}\ \bibnamefont
  {Braginsky}}\ and\ \bibinfo {author} {\bibfnamefont {F.}~\bibnamefont
  {Khalilli}},\ }\href
  {http://www.cambridge.org/us/academic/subjects/physics/quantum-physics-quant%
um-information-and-quantum-computation/quantum-measurement} {\emph {\bibinfo
  {title} {{Quantum Measurement}}}}\ (\bibinfo  {publisher} {Cambridge
  University Press},\ \bibinfo {year} {1992})\BibitemShut {NoStop}%
\bibitem [{\citenamefont {{A. Buonanno}}\ and\ \citenamefont {{Y.
  Chen}}(2002)}]{BuCh2002s}%
  \BibitemOpen
  \bibfield  {author} {\bibinfo {author} {\bibnamefont {{A. Buonanno}}}\ and\
  \bibinfo {author} {\bibnamefont {{Y. Chen}}},\ }\href
  {http://journals.aps.org/prd/abstract/10.1103/PhysRevD.65.042001} {\bibfield
  {journal} {\bibinfo  {journal} {Phys. Rev. D}\ }\textbf {\bibinfo {volume}
  {65}},\ \bibinfo {pages} {042001} (\bibinfo {year} {2002})}\BibitemShut
  {NoStop}%
\bibitem [{\citenamefont {Clerk}\ \emph {et~al.}(2010)\citenamefont {Clerk},
  \citenamefont {Devoret}, \citenamefont {Girvin}, \citenamefont {Marquardt},\
  and\ \citenamefont {Schoelkopf}}]{Clerk2008s}%
  \BibitemOpen
  \bibfield  {author} {\bibinfo {author} {\bibfnamefont {A.~A.}\ \bibnamefont
  {Clerk}}, \bibinfo {author} {\bibfnamefont {M.~H.}\ \bibnamefont {Devoret}},
  \bibinfo {author} {\bibfnamefont {S.~M.}\ \bibnamefont {Girvin}}, \bibinfo
  {author} {\bibfnamefont {F.}~\bibnamefont {Marquardt}}, \ and\ \bibinfo
  {author} {\bibfnamefont {R.~J.}\ \bibnamefont {Schoelkopf}},\ }\href
  {\doibase 10.1103/RevModPhys.82.1155} {\bibfield  {journal} {\bibinfo
  {journal} {Rev. Mod. Phys.}\ }\textbf {\bibinfo {volume} {82}},\ \bibinfo
  {pages} {1155} (\bibinfo {year} {2010})}\BibitemShut {NoStop}%
\bibitem [{\citenamefont {Gardiner}\ and\ \citenamefont
  {Collett}(1985)}]{Gardiner1985s}%
  \BibitemOpen
  \bibfield  {author} {\bibinfo {author} {\bibfnamefont {C.~W.}\ \bibnamefont
  {Gardiner}}\ and\ \bibinfo {author} {\bibfnamefont {M.~J.}\ \bibnamefont
  {Collett}},\ }\href {\doibase 10.1103/PhysRevA.31.3761} {\bibfield  {journal}
  {\bibinfo  {journal} {Phys. Rev. A}\ }\textbf {\bibinfo {volume} {31}},\
  \bibinfo {pages} {3761} (\bibinfo {year} {1985})}\BibitemShut {NoStop}%
\bibitem [{\citenamefont {Blow}\ \emph {et~al.}(1990)\citenamefont {Blow},
  \citenamefont {Loudon}, \citenamefont {Phoenix},\ and\ \citenamefont
  {Shepherd}}]{Blow1990s}%
  \BibitemOpen
  \bibfield  {author} {\bibinfo {author} {\bibfnamefont {K.~J.}\ \bibnamefont
  {Blow}}, \bibinfo {author} {\bibfnamefont {R.}~\bibnamefont {Loudon}},
  \bibinfo {author} {\bibfnamefont {S.~J.~D.}\ \bibnamefont {Phoenix}}, \ and\
  \bibinfo {author} {\bibfnamefont {T.~J.}\ \bibnamefont {Shepherd}},\ }\href
  {\doibase 10.1103/PhysRevA.42.4102} {\bibfield  {journal} {\bibinfo
  {journal} {Phys. Rev. A}\ }\textbf {\bibinfo {volume} {42}},\ \bibinfo
  {pages} {4102} (\bibinfo {year} {1990})}\BibitemShut {NoStop}%
\end{thebibliography}
\end{document}